\documentclass[iop]{emulateapj}
\usepackage{natbib}

\usepackage{graphicx}

\submitted{Received: 2011 November 30; Accepted: 2011 December 1}
\journalinfo{\sc Accepted by \emph{Publications of the Astronomical Society of the Pacific}}

\shortauthors{Bell et al.}
\shorttitle{H$\alpha$ Emission Variability in Active M Dwarfs}

\begin{document}

\title{H$\alpha$ Emission Variability in Active M Dwarfs}

\author{Keaton J. Bell\footnotemark[1,2,*]}
\author{Eric J. Hilton\footnotemark[1,3]}
\author{James R. A. Davenport\footnotemark[1]}
\author{Suzanne L. Hawley\footnotemark[1]}
\author{Andrew A. West\footnotemark[4]}
\author{Allen B. Rogel\footnotemark[5]}

\footnotetext[1]{Astronomy Department, University of Washington, Seattle, WA 98195}
\footnotetext[2]{Department of Astronomy, University of Texas, Austin, TX 78712}
\footnotetext[3]{Dept. of Geology and Geophysics and Institute for Astronomy, University of Hawaii at Manoa, Honolulu, HI 96822}
\footnotetext[4]{Department of Astronomy, Boston University, Boston, MA 02215}
\footnotetext[5]{Department of Physics \& Astronomy, Bowling Green State University, Bowling Green, OH 43403}
\footnotetext[*]{Direct correspondence to: \href{mailto:keatonb@astro.as.utexas.edu}{keatonb@astro.as.utexas.edu}}

\begin{abstract}

  We use $\sim$12,000
  spectra of $\sim$3,500 magnetically active M0-M9 dwarfs from the Sloan Digital
  Sky Survey taken at 10-15 minute intervals, together with
  $\sim$300 spectra of $\sim$60 M0-M8 stars obtained hourly with the
  Hydra multi-object spectrometer, to probe H$\alpha$ variability on
  timescales of minutes to weeks.  With multiple observations for
  every star examined, we are able to characterize fluctuations in
  H$\alpha$ emission as a function of activity strength and
  spectral type.  
  Stars with greater magnetic activity (as quantified by
  L$_{\rm{H}\alpha}$/L$_{bol}$) are found to be less variable at all spectral types.  
  We attribute this result to the stronger level of persistent emission in the high activity 
  stars, requiring a larger heating event  in order to produce measurable variability. 
  We also construct H$\alpha$ structure functions to
  constrain the timescale of variability. The more active objects with lower
   variability exhibit a characteristic timescale longer than an hour, likely 
   due to larger, longer lasting heating events, while the less active 
   objects with higher variability have a characteristic timescale shorter than
  15 minutes.

\end{abstract}

\keywords{stars: activity --- stars: chromospheres --- stars: flare --- stars: low-mass --- stars: magnetic fields}

\section{Introduction}
\label{sec:intro}

M dwarfs, the least massive and most numerous stars in our Galaxy,
provide remote laboratories for observing high-energy magnetic
phenomena in stellar atmospheres.  The red and dim nature of M dwarfs
makes the H$\alpha$ spectral feature the most accessible tracer of
magnetic activity in their convective outer layers 
\citep{Bopp1978,Walkowicz2009}.  M dwarfs with H$\alpha$ in emission are
classified as magnetically active, with a greater fraction of stars
showing H$\alpha$ emission at later spectral types
\citep{Hawley1996,Gizis2000,West2011}.  Magnetic activity is
correlated with energetic flaring events that last seconds to hours
\citep{moffett74,kowalski2010} and photometric brightness increases
as large as 8.2 magnitudes in the \emph{B}-band \citep{Almeida2011}.
The incidence of H$\alpha$ emission and large flares decreases for
stars farther from the Galactic Plane
\citep{West2006,West2008,kowalski2009,Hilton2010, 2011AJ....141...50W}.  This is interpreted as an age
effect, leading to activity timescales of 1--8 billion years for
early--late M dwarfs respectively.

Variability in quiescent (non-flaring) H$\alpha$ emission has been
attributed to both stellar rotation and M dwarf activity cycles.
\citet{Bopp1978} related sequential H$\alpha$ measurements to orbital 
phase and suggested that the observed
changes were caused by the rotation of localized active regions
in and out of view on timescales of
days.  Activity cycles with periods of a few years have been
identified for M dwarfs by \citet{Cincunegui2007b} and
\citet{Buccino2011}, and may affect the observed strength of H$\alpha$
emission.  Although large flares are relatively infrequent events
\citep{Hilton2011}, sporadic low-level magnetic field reconnection
events (``microflares'') may also contribute to inherent H$\alpha$ emission
variations in active M dwarfs.

There have been some recent investigations of H$\alpha$ variability on
timescales of minutes to days.  \citet{Gizis2002} measured the
H$\alpha$ equivalent widths (EWHA) for a sample of more than 600
active M0--M6 dwarfs from the Palomar/Michigan State University (PMSU) 
survey of nearby stars \citep{reid1995}, and
found that the variation in
repeated measurements was near 15\% 
for stars with mean EWHA,  
$\langle$EWHA$\rangle$
$<$ 5\AA, while
the variability for stars with 
$\langle$EWHA$\rangle$ $>$ 5\AA\ 
was in excess of 30\%. 
In their study of EWHA variations on later type stars, 
\citet{Lee2010} identified
significant variability in $\sim$80\% of their sample of 
43 M4--M8 stars.  They found that the majority of
variability events had timescales of 30 minutes or longer,
and that the later-type stars were more variable,
as measured by $R$(EW) = max(EWHA)/min(EWHA).
Similarly,
\citet{Kruse2010} found an increase in the same
metric (which they refer to as $R_{EW}$) at later spectral types for
timescales from 15 minutes to an hour in the 
Sloan Digital Sky Survey \citep[SDSS;][]{york2000}
Data Release 7 \citep[DR7;][]{dr7} component spectra.  They also 
noted that $R_{EW}$ was larger for intermittent sources.

Our analysis uses a set of spectra with exposure times of $\sim$1 hour
from the multi-object Hydra spectrograph on the WIYN telescope at Kitt
Peak National Observatory, as well as the component
spectra from SDSS, to extend the evaluation of H$\alpha$ variability
using new metrics and a broader range of timescales.  We 
employ a construct known as the structure function, which is
commonly used in quasar variability analyses
\citep[e.g.][]{Simonetti1985,MacLeod2011}, to further constrain the timescales
of H$\alpha$ emission variability.

The spectroscopic M dwarf samples we used are detailed in
Section 2, along with our methods for determining 
spectral types and measuring H$\alpha$ emission strength. 
Analysis of H$\alpha$ emission variability with spectral type, 
activity strength and timescale 
is presented in
Section 3, and we summarize our findings in Section 4.

\section{Data}
\label{sec:data}
\begin{figure}[t]
\centering
  \scalebox{0.37}{\includegraphics*[0.5in,0in][9.5in,2.853in]{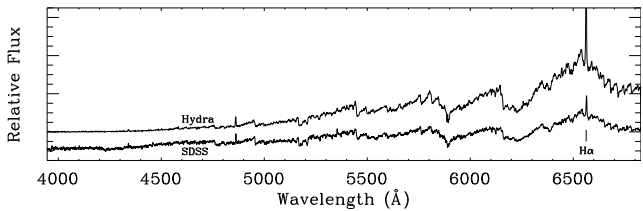}}
  \caption{Sample high-S/N active M4 dwarf spectra from the Hydra (top) and 
SDSS (bottom) data are shown.  The data
  have moderate spectral resolution (2000-4000) and wavelength coverage
from $\sim$4000-9000\AA\ for SDSS and $\sim$4000-7000\AA\ for Hydra.}
  \label{spectra}
\end{figure}

\begin{figure}[t]
\centering
  \scalebox{0.53}{\includegraphics*[0.15in,0.19in][6.6in,4.56in]{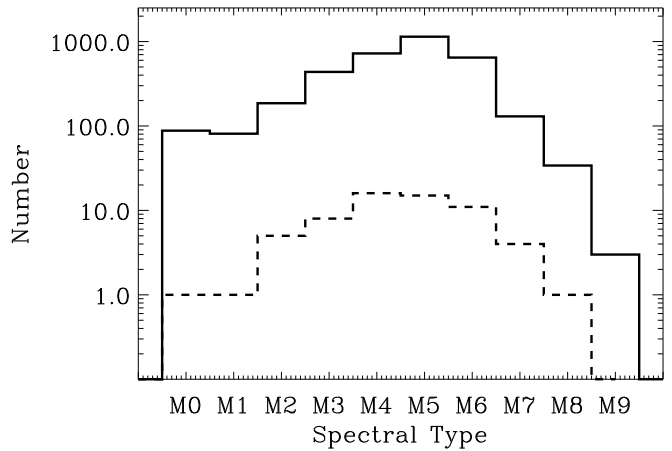}}
  \caption{The distribution of spectral types is similar 
for the magnetically active M dwarfs in the SDSS (solid line) and Hydra (dashed line) samples.}
  \label{sptype}
\end{figure}

Our data come from two sources that are discussed in detail below: the
Hydra instrument on the WIYN telescope (Section 2.1), and the Sloan
Digital Sky Survey (Section 2.2).  Representative spectra from each
sample are displayed in Figure \ref{spectra}.
The spectral type distributions of the
active M dwarfs in the two samples are shown in Figure \ref{sptype}.  These
moderate-resolution data sources (Hydra: R$\sim$4000, SDSS: R$\sim$2000) 
are useful for our H$\alpha$
variability analysis because they each provide multiple spectra of the
same objects at well-defined time intervals.

\subsection{WIYN Hydra Data}

Our analysis employs data that were taken over 27 nights between
January 2002 and July 2008 as part of the {\it Chandra}
Multiwavelength Plane Survey \citep[ChaMPlane;][]{Grindlay2005}, which
was designed to investigate the fraction of diffuse
X-ray emission in the Galactic plane due to unresolved
X-ray point sources.  
Candidate point sources were identified using $VRI$H$\alpha$ photometry
obtained with the Mosaic cameras at the Cerro Tololo Inter-American Observatory and 
the Kitt Peak National Observatory 4-meter telescopes. 
Spectra were obtained with the Hydra
multi-object fiber-fed spectrograph on the 3.5-meter WIYN telescope for
objects with H$\alpha-R< -0.3$,
indicating excess H$\alpha$ emission, as well as optically detected
objects with corresponding X-ray emission in archival Chandra images. Although the survey was designed to identify accretion powered sources such as cataclysmic variables and low-mass X-ray binaries, 
initial analysis by the ChaMPlane team
\citep{Rogel2006} revealed a sizable population of M dwarfs in the sample,
whose spectra and photometry we utilize in this study.

We assigned spectral types to the Hydra data using the Hammer software
\citep{Covey2007} in manual mode, because the
flux calibration and signal-to-noise ratio (typically $\sim$10) of the spectra, especially at blue 
wavelengths, did not allow automatic typing.  Each object 
was visually 
compared to the Hammer set of M dwarf templates \citep{Bochanski2007}. 
The resulting spectral types are accurate to within 1--2 types. 
We verified that the spectral types were broadly consistent with the measured $V-I$ colors
of the objects.

\begin{figure}[b]
\centering
  \scalebox{0.57}{\includegraphics*[0.67in,0.17in][6.7in,4.66in]{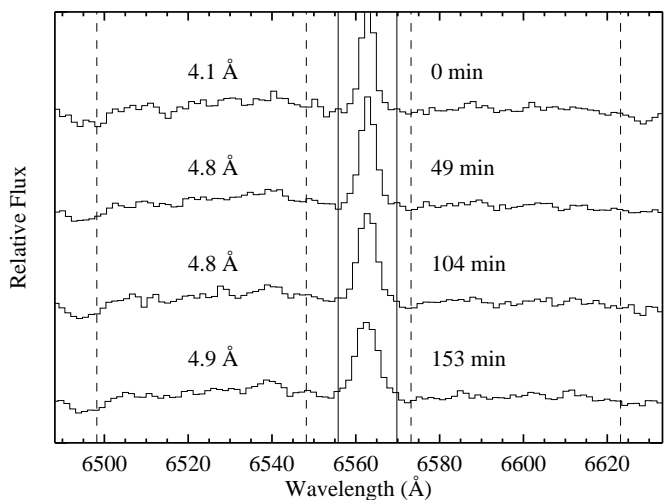}}
  \caption{H$\alpha$ line profiles in the Hydra data for repeated 
observations of an M4 dwarf.  The solid
vertical lines indicate the boundaries of the line region used in the 
equivalent width calculations, 
and the dashed lines mark the pseudo-continuum regions on either side.  
The measured equivalent width for each
spectrum is indicated to the left of the emission line, and the time elapsed 
since the beginning of the 
first exposure is shown on the right.}
\label{sampleewha}
\end{figure}

The strength of H$\alpha$ emission in each spectrum was quantified by
measuring the equivalent width of the line.  We followed the procedure
described in \citet{West2004} and adopted their definitions for the H$\alpha$
line and continuum regions. Figure \ref{sampleewha}
identifies these regions for the four Hydra spectra obtained for a
representative M4 dwarf. We visually inspected every spectrum in the
sample to ensure that the EWHA measurements were not contaminated
by cosmic rays or other artificial defects.

Because the equivalent width is measured with respect to the local
continuum level, it can be misleading to compare EWHA among stars
of different spectral type, and hence continuum flux near H$\alpha$.
We therefore computed the ratio of 
the H$\alpha$ line luminosity to the bolometric
luminosity (L$_{\rm{H}\alpha}$/L$_{bol}$) for each object using the
$\chi$ conversion factors of \citet{West2008X}.
This quantity represents the continuum-independent activity level,
which allows for H$\alpha$ emission to
be compared between stars of different spectral types.

Our analysis was limited to the chromospherically active M dwarfs,
identified by the criteria of \citet{West2011} that define detectable
H$\alpha$ emission.
Because we intend to use these data in a
time-domain analysis, we further restricted the sample
to include only those stars with at least three spectra
that were identified as active. Our final Hydra sample contains 312 
spectra from 62 active M0--M8 dwarfs, and allows us to probe H$\alpha$
emission variability on timescales from one to three hours.

\subsection{SDSS Spectra}

All SDSS spectra are composites of several (typically 3-5) spectra
with $\sim$10-15 minute exposure times.  The individual component spectra 
(with typical signal-to-noise ratio $\sim$7)
were initially made public
as part of the SDSS Data Release 6 \citep[DR6;][]{DR6}.
Previous studies have examined the spectral variability of M dwarfs
using the individual SDSS spectra from DR6 \citep{Hilton2010} and 
DR7 \citep{Kruse2010}; our analysis expands on these efforts.  
We use the \citet{Hilton2010} sample, which has the spectral type
for each object assigned from the composite spectra presented 
in \citet{West2008}.  The EWHA measurements for the component spectra 
followed the same prescription as for the Hydra data.
L$_{\rm{H}\alpha}$/L$_{bol}$ values were obtained from
the EWHA calculated for each object, and we verified that they
agreed with those measured from the
composite spectra given in \citet{West2008}.
We further restricted our sample to include only the 
persistently active objects
that showed measurable H$\alpha$ emission in every
individual exposure.  We note that this reduced the 
\cite{Hilton2010} active sample by less than 3\%.

Our final SDSS sample consists of 11,989 spectra for
3,469 active M0--M9 dwarfs, and allows us to examine variability on
timescales from $\sim$15--100 minutes.  Nearly 700 stars were observed on
multiple days, allowing us to further probe H$\alpha$
emission changes on timescales of $\sim$1-20 days.

\section{Variability Analysis and Results}
\label{sec:results}
We first consider H$\alpha$ variability as a function of spectral type, 
comparing our results with variability indicators used in previous 
analyses (Section 3.1). Our preferred metric for variability is 
$\sigma_{EWHA}/\langle$EWHA$\rangle$, and we investigate this metric as a 
function of activity strength (L$_{\rm{H}\alpha}/$L$_{bol}$; Section 3.2). Finally, we compute 
structure functions to investigate characteristic
timescales in the low- and high-variability
subsets of our data (Section 3.3).

\subsection{H$\alpha$ Variability with Spectral Type}

\begin{figure}[t]
\centering
  \scalebox{0.55}{\includegraphics*[0.5in,0.18in][6.64in,5.6in]{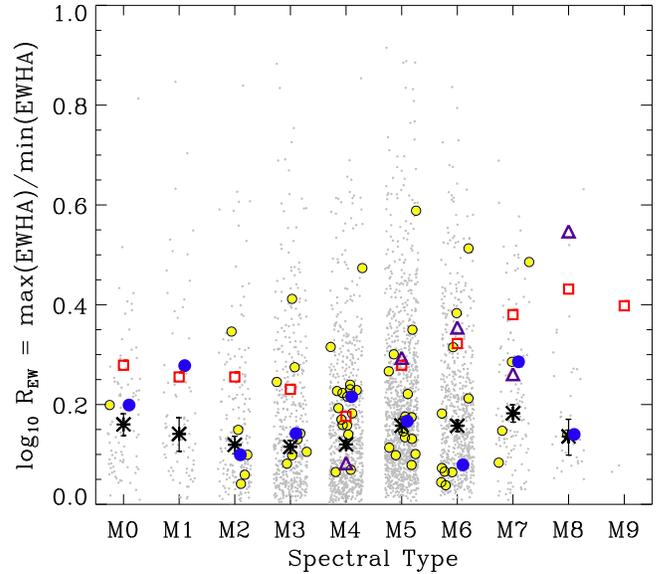}}
  \caption{The $R_{EW}$ variability metric is shown as a function of spectral
type for the SDSS data (light gray dots) and Hydra data (light blue diamonds),
illustrating distributions with large numbers of low-variability objects and
extended tails toward higher variability that are more noticeable at later
types.  The distributions are artificially broadened at each spectral type to more clearly show the actual number of objects. 
The median of each type is given by the black asterisks (SDSS) and blue
filled circles (Hydra).  There is no obvious trend in the medians with spectral
type.  The red squares and purple triangles denoting results from \cite{Kruse2010} and \cite{Lee2010} respectively, are shown for comparison.  See text for discussion.}
\label{maxmin}
\end{figure}

\citet{Lee2010} and \citet{Kruse2010} both found that 
H$\alpha$ variability increases at later spectral type, using
the $R_{EW}$ metric.  Figure \ref{maxmin}
illustrates $R_{EW}$
as a function of spectral type for our Hydra and SDSS samples.
In agreement
with their analyses, we find that the distributions
are not symmetric, but have significant extensions 
to large values of $R_{EW}$, 
with these extensions being more pronounced at later spectral type.
However, in our samples the median values (the median is preferred 
as a statistic instead of the mean for these skewed distributions) 
scatter around a roughly constant value.
This is in contrast to the \citet{Kruse2010} and \citet{Lee2010} results, 
shown as red squares and purple triangles
respectively.  
We attribute the difference to the criteria used to identify the active stars.
\citet{Kruse2010} and \citet{Lee2010} include stars that they characterize as ``intermittently''
variable (i.e. only showing H$\alpha$ emission in a subset of their spectra), and they exclude weakly active stars, which they classify as inactive.
These differ from our analysis as a result of their non-uniform equivalent width measurement criteria.
Their intermittently variable stars are more prevalent at later types and systematically exhibit larger
values of the $R_{EW}$ metric.  We are able to confidently identify H$\alpha$ at much 
smaller equivalent width, even in late--type M dwarfs, provided the spectra have
sufficiently high signal-to-noise ratios \citep{West2008}. Evidently, including these less active, but persistent, objects leads to the approximately flat distribution outlined by the SDSS and Hydra medians in Figure \ref{maxmin}.

\subsection{H$\alpha$ Variability with Activity Strength}

\begin{figure}[t]
\centering
  \scalebox{0.52}{\includegraphics*[0.32in,0.60in][6.65in,8.0in]{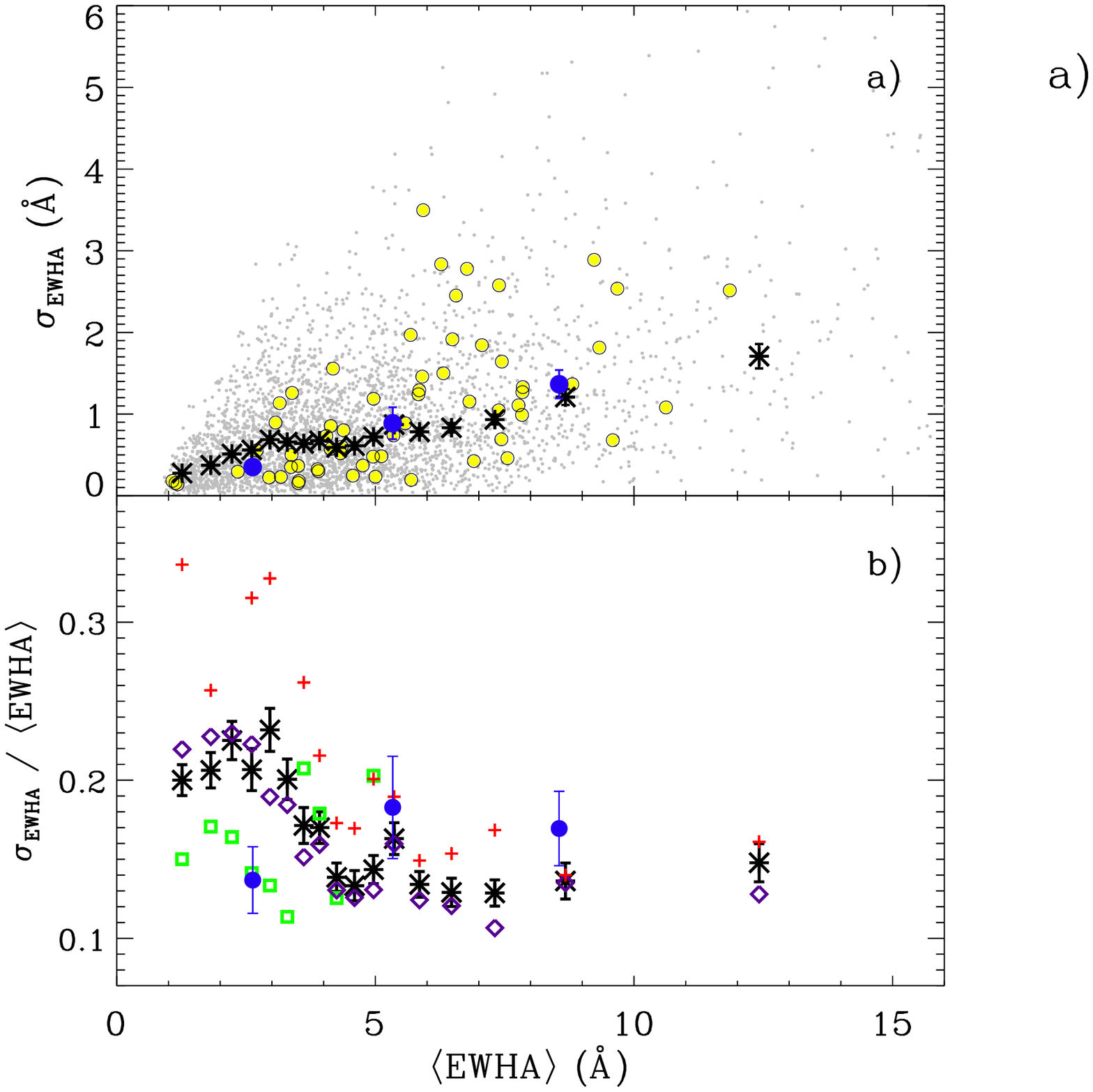}}
  \caption{
a) The standard deviation in EWHA ($\sigma_{EWHA}$) versus 
mean EWHA 
($\langle$EWHA$\rangle$) 
is shown for each star in the Hydra 
and SDSS samples. 
Medians of representative bins are also shown (black asterisks for SDSS, blue circles for Hydra).  There is a rising trend toward larger $\sigma$ and higher scatter as
$\langle$EWHA$\rangle$ increases.
b) 
The medians of the binned data for the fractional variability metric 
$\sigma_{EWHA}$/$\langle$EWHA$\rangle$ are shown grouped by spectral
type for the SDSS data (M0--M2 as green squares, M3--M5 as purple diamonds 
and M6--M9 as red crosses).  
There is a clear separation by spectral type,
and a general decreasing trend toward lower variability at larger
$\langle$EWHA$\rangle$.
The median SDSS and Hydra data have the same symbols as in the top panel.
}
\label{sigmaewha}
\end{figure}

\citet{Gizis2002} used the standard deviation of repeated EWHA 
measurements ($\sigma_{EWHA}$) for each star as a metric for line 
strength variability, and found that this value increased with
$\langle$EWHA$\rangle$ for their sample. We investigate that 
result in Figure \ref{sigmaewha}a, 
which illustrates the variation of $\sigma_{EWHA}$ with 
$\langle$EWHA$\rangle$ for stars in both the Hydra and SDSS samples. 
The median $\sigma_{EWHA}$ for bins containing roughly equal numbers of 
stars are also shown.  The abrupt cutoff at 
$\langle$EWHA$\rangle\sim1$\AA\ is due to our requirements for
measurable H$\alpha$ equivalent width in these data \citep{West2008}.

We see that the median
$\sigma_{EWHA}$ is near zero for low $\langle$EWHA$\rangle$, rising
to $\sim$0.8\AA\ at $\langle$EWHA$\rangle \sim$ 3--6\AA\, and
increasing to $\sim$1.8\AA\ at 
$\langle$EWHA$\rangle>$ 10\AA.  
Our well-sampled data show a smooth, continuously increasing scatter of $\sigma_{EWHA}$ with
$\langle$EWHA$\rangle$ for the median points, in qualitative agreement with the much smaller sample of \citet{Gizis2000}.
We also note that the maximum value of $\sigma_{EWHA}$ increases with $\langle$EWHA$\rangle$.
%We do not, however, observe an abrupt increase in the scatter near
%$\langle$EWHA$\rangle\sim5$\AA, which they claimed was evident in their data.

As noted above, the absolute equivalent width values depend on spectral 
type due to the changing continuum flux near H$\alpha$, with
earlier type stars having relatively smaller equivalent width due to their
higher continuum flux.  We illustrate this effect in  
Figure \ref{sigmaewha}b using the fractional
variability metric 
$\sigma_{EWHA}/\langle$EWHA$\rangle$.  
The skewed nature of the $\sigma$ distributions at a given
$\langle$EWHA$\rangle$,
evident in Figure \ref{sigmaewha}a (and analogous to the skewed distributions
in $R_{EW}$ shown in Figure \ref{maxmin}), again leads us to use the
median value to characterize
$\sigma_{EWHA}/\langle$EWHA$\rangle$
for each $\langle$EWHA$\rangle$ bin.

The spectral type dependence of $\langle$EWHA$\rangle$ is clearly indicated 
by the separation of the early (M0--M2, green squares), mid (M3--M5, purple
diamonds) and late (M6--M9, red crosses) types for the SDSS sample.  
The median 
$\sigma_{EWHA}/\langle$EWHA$\rangle$ values
for all spectral types are shown as black asterisks (SDSS) and blue filled
circles (Hydra).
The variability metric decreases with increasing
$\langle$EWHA$\rangle$ from $\sim$0.2 for stars 
with low EWHA to $\sim$0.14 for stars with $\langle$EWHA$\rangle>5$\AA.

\begin{figure}[t]
  \scalebox{0.56}{\includegraphics*[0.55in,0.20in][6.6in,5.644in]{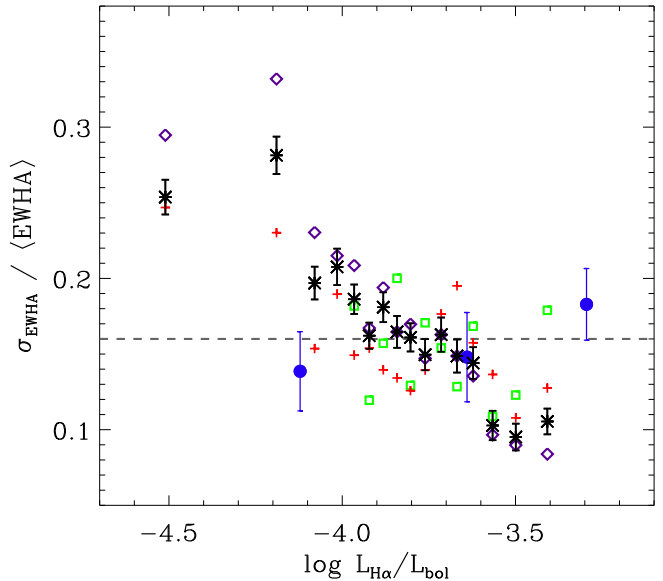}}
  \caption{The same medians of the fractional variability metric
($\sigma_{EWHA}$/$\langle$EWHA$\rangle$) from Figure \ref{sigmaewha}b
are shown as a function of activity strength
log L$_{\rm{H}\alpha}$/L$_{bol}$.  The symbols are the
same as in Figure \ref{sigmaewha}b. 
The downward trend in the SDSS data shows that variability decreases in more active M dwarfs,
regardless of spectral type. The Hydra data do not show a coherent trend,
which we attribute to poor sampling and a bias toward very active, early
type stars.}
\label{lhalbol}
\end{figure}

A better measure of magnetic activity strength that accounts for the 
continuum dependence on spectral type is 
L$_{\rm{H}\alpha}$/L$_{bol}$.
Figure \ref{lhalbol} illustrates the same 
fractional variability metric 
$\sigma_{EWHA}/\langle$EWHA$\rangle$ as in 
Figure \ref{sigmaewha}b,
versus activity strength.
The spectral types are no longer separated; in fact the three spectral
type groups from Figure 
\ref{sigmaewha}b,
have shifted horizontally so that they now overlap and form 
a smooth distribution.  It is clear that
the underlying variability relationship is not a function of spectral type but
rather of activity strength.
However, the median fractional variability (black asterisks) {\it decreases} with
increasing activity from
$\sim$0.3 at log(L$_{\rm{H}\alpha}$/L$_{bol})\sim-4.5$
to $\sim$0.1 at $\log($L$_{\rm{H}\alpha}$/L$_{bol})
\sim-3.5$.  The median value for the variability metric across the
entire sample is 0.16, shown as the dashed line on the Figure.

The Hydra data (blue filled circles) in Figures 
\ref{sigmaewha}b and \ref{lhalbol} show relatively flat distributions
that are influenced by small numbers, uneven spectral type sampling,
and a bias toward very active stars especially at early spectral types
(due to the H$\alpha-R$ color selection of the targets).  Therefore, they
are not as useful as the SDSS data for understanding the behavior of the
variability metric with activity strength.

\begin{figure*}[ht]
\centering
\scalebox{0.77}{\includegraphics*[0.56in,0.053in][9.51in,3.27in]{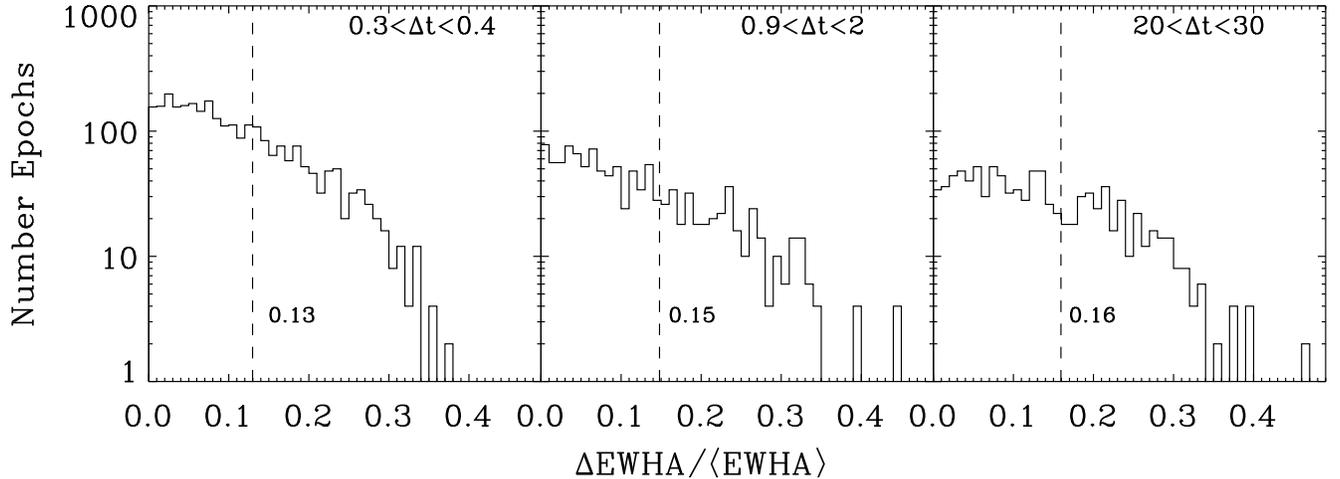}}
  \caption{Three time bins from the structure function of the low-variability 
SDSS sources.
   Each point on the structure function in Figure \ref{strfunc} is a 
normalized RMS (denoted by the vertical dashed lines)
  calculated from a distribution of $\Delta$EWHA/$\langle$EWHA$\rangle$ that is 
  populated by pairs of observations separated by a specific range of 
  time, $\Delta$t (given in hours at the top of each panel).}
  \label{strfunchist}
\end{figure*}

The picture we have developed is that highly active stars of all spectral
types are relatively less variable than low activity stars.  This only
becomes evident when the well-known
underlying relationship between activity strength and spectral type is
removed, as in Figure \ref{lhalbol}.   Physically, 
the strong level of persistent emission in the high activity stars 
may be due to a large
surface coverage of active regions, while the low activity stars may
have a relatively smaller filling factor.  An increase in H$\alpha$ emission, 
for example due to a new active
region appearing on the surface, or a magnetic heating episode or microflare in
one of the existing active regions, will then manifest as barely visible in the more active star, 
while it will contribute significant new emission in the less active star.
This would naturally explain why the less active stars are relatively more variable.

\subsection{Structure Functions and Timescales of H$\alpha$ Variability}

Structure functions are often used to explore the
variability timescales present in time-resolved observations of
quasars \citep[e.g.][]{Simonetti1985,MacLeod2011}.  They represent the
change observed between two measurements of some variable 
quantity, typically broadband flux in the case of quasars, 
as a function of the time between measurements ($\Delta$t).  
It is expected that if the
observed variability operates on a characteristic timescale
($\tau$), then for $\Delta$t $<$ $\tau$, the measurements will be 
correlated. The variability amplitude
will increase with greater time separation as the observations
sample more of the total variability range until 
$\Delta$t = $\tau$.  For $\Delta$t $>$ $\tau$,
the structure function maintains a constant amplitude as the measurements 
are uncorrelated and randomly sample the full variability range.

We employ structure functions here to quantify the typical change in the 
equivalent width measurements for
active M dwarf spectra in many bins of time separation, $\Delta$t, and to constrain the characteristic timescale of variability, $\tau$. 
We divide our samples at
the median value of the fractional variability metric in 
Figure \ref{lhalbol} into the less active but
more-variable stars
($\sigma_{EWHA}$/$\langle$EWHA$\rangle$ $>$ 0.16), 
and the more active but less-variable
stars ($\sigma_{EWHA}$/$\langle$EWHA$\rangle$ $<$ 0.16). 
We then calculate the change in H$\alpha$
emission as a fraction of the mean value,
$\Delta$EWHA/$\langle$EWHA$\rangle$, between every pair of 
repeated measurements for the same M dwarf. 

To illustrate how the structure function amplitude is calculated, the 
distributions for three SDSS time bins from the less-variable 
(more active) subsample are presented in 
Figure \ref{strfunchist}.  
Each time bin (the $\Delta$t range indicated on the Figure) contributes one
point on the structure function. The typical change between 
measurements for a given $\Delta$t bin is
taken as the RMS of the $\Delta$EWHA/$\langle$EWHA$\rangle$ distribution, 
normalized by the number of samples in that time bin. This accounts 
for the uneven sampling of time separations in our datasets. The 
normalized RMS values for these three bins increase from
0.13 to 0.16 with increasing $\Delta$t. 
The SDSS structure functions were constructed from similar normalized 
RMS values that were calculated for 14 time bins
spanning the entire range of $\Delta$t sampled by our data.
The time bins were chosen to provide good
statistics at short $\Delta$t intervals (15 minutes to 2 hours), 
where there are many observations, and to cluster around typical 
$\Delta$t values for longer intervals ($e.g.$ one day, several days).  
The Hydra structure functions were only computed
for a few time intervals, reflecting the approximately hourly cadence
of the observations.

\begin{figure}[t]
\centering
\scalebox{0.5}{\includegraphics*[0.05in,0.31in][6.94in,4.67in]{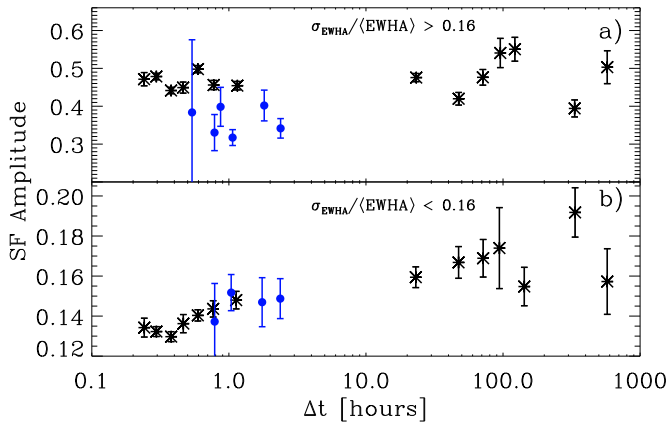}}
  \caption{The structure functions for a) the high-variability 
(less active) stars;
and b) the low-variability (more active) stars, for the SDSS (black asterisks) and
Hydra (blue filled circles) samples.  The low-variability stars show an
increasing structure function amplitude between 0.4-1 hour, 
indicating a timescale of more than an hour for the
variability.  The high-variability stars apparently vary on 
timescales shorter than our smallest time separation ($\sim$15 minutes).  
The Hydra data are discussed separately in the text.}
  \label{strfunc}
\end{figure}

Figure \ref{strfunc}a illustrates the SDSS (black asterisks) and 
Hydra (blue filled circles) structure functions for the 
high-variability sample 
($\sigma_{EWHA}$/$\langle$EWHA$\rangle >$ 0.16), 
and Figure \ref{strfunc}b gives the structure
functions for the low-variability sample 
($\sigma_{EWHA}$/$\langle$EWHA$\rangle <$ 0.16). 
The timescales range
from 15 minutes to 20 days. The irregular binning in 
$\Delta$t reflects the non-uniform separations of the measurements in 
time due to the sampling cadences of the surveys. We also require
that the shortest time bins ($\Delta$t $<$ 2 hours)
in the SDSS structure function contain at least 
150 measurement pairs, in order to obtain well-defined normalized RMS values
(see Figure \ref{strfunchist}).  At larger time separations, the sample
sizes are smaller and the error bars on the structure function amplitudes
are correspondingly larger.

The low-variability structure function for the SDSS sample in Figure \ref{strfunc}b shows 
a significant trend with time, indicating
increasing variability up to a timescale of at least an hour. 
The plateau at timescales shorter than $\Delta$t $\sim$ 0.4 hours suggests 
that measurement uncertainties may dominate any variability amplitude effects 
below that time separation. 

The high-variability structure function (Figure \ref{strfunc}a), on the other hand, 
shows a flat distribution at all times for the SDSS sample.  The structure
function amplitude is much larger, which is indicative of the higher
level of variability seen in this sample.
The measurement uncertainties are much less than the detected RMS (structure function amplitude), 
indicating that the variability timescale is shorter than our smallest time
separation bin, $\sim$15 minutes.

While sparsely sampled, the Hydra data reinforce this interpretation. 
In Figure \ref{strfunc}a, both the SDSS and Hydra exposure times are 
longer than the underlying characteristic timescale for the 
high-variability data, and therefore all measurement 
pairs are randomly sampling the intrinsic equivalent width 
variations, leading to flat structure function amplitudes.
The longer exposure time for the Hydra data (an hour compared to 
10-15 minutes for SDSS) 
damps the stochastic 
variations between exposures, and the amplitude of the Hydra structure 
function is correspondingly smaller.  The low-variability 
Hydra sample in Figure \ref{strfunc}b has a characteristic timescale 
that is apparently comparable to the exposure time, and thus the 
structure function amplitude is flat over the 1--3 hour time
separations that are sampled.
In this case, the structure function amplitude is consistent between
the SDSS and Hydra samples.

The increase in variability amplitude with time seen in our
low-variability (high activity) sample is reminiscent of the \citet{Lee2010}
result that there was a much larger number of H$\alpha$ 
emission events that lasted at least 30 minutes compared to the number of 
shorter events they observed.  
Their sample was primarily composed 
of very active stars which is comparable to this subset of our data.

Following the discussion in \S3.2, we interpret the timescale results
as follows.  The low-variability (high activity) stars are covered
with spots, so that small variations are not easily visible against
the high background.  In order to see a noticeable variation, a rather
significant event must occur.  Since higher-energy events are known
to last longer \citep[e.g.][]{lme,hp1991}, this leads to a longer timescale 
for events that are
strong enough to be detected above the persistent activity.
For the high-variability (low activity) stars, small events will
cause a noticeable change, and these occur on short timescales, below
the $\sim$15 minute limit of our sampling cadence.

\section{Summary}
\label{sec:discussion}

We used two spectroscopic samples of active M dwarfs to examine
H$\alpha$ emission variability on timescales from minutes to weeks.

We found that the H$\alpha$ variability measured using the $R_{EW}$
metric remains relatively
constant as a function of spectral type, in contrast to previous results.
However, using our fractional variability metric 
($\sigma_{EWHA}$/$\langle$EWHA$\rangle$),
we showed that the higher-variability stars have relatively low
activity strength  (L$_{\rm{H}\alpha}$/L$_{bol}$), and conversely
the low-variability stars have high activity strength.  
This naturally leads to an apparent dependence on spectral
type because the later type stars typically have lower activity strength. We speculated
that the physical reason for the higher variability in the low activity stars is that small changes
in H$\alpha$ emission, e.g. due to small-scale flaring or the appearance of a new
active region on the stellar surface, would have a relatively larger
effect on the less active stars.

We investigated the timescales of H$\alpha$ emission variability
using structure functions.  The observed H$\alpha$
variations for low-variability stars 
($\sigma_{EWHA}$/$\langle$EWHA$\rangle<$ 0.16) 
occur on a timescale longer than an hour,
while the high-variability M dwarfs exhibit a timescale
that is shorter than our sampling time of $\sim$15 minutes.  
Neither the low nor high variability samples show significant
changes in the structure function on timescales longer than a day.
We speculated that the low-variability sample, which is mainly composed
of the higher activity strength stars, requires more energetic and
hence longer-lived emission events in order to be detected above the
strong persistent surface activity, leading to a longer characteristic
variability timescale.

Better time resolution in spectroscopic M dwarf monitoring, as well as
a wider and better-sampled range of time separations,
will allow for a more accurate determination of the timescales of
H$\alpha$ variability in low-mass dwarfs and may lead to a
better understanding of the physical mechanisms that cause these
changes.

We acknowledge NSF grant AST 08-07205, and 
thank David Schlegel for help obtaining the individual component SDSS spectra.

Funding for the SDSS and SDSS-II has been provided by the Alfred
P. Sloan Foundation, the Participating Institutions, the National
Science Foundation, the U.S. Department of Energy, the National
Aeronautics and Space Administration, the Japanese Monbukagakusho, the
Max Planck Society, and the Higher Education Funding Council for
England.  The SDSS Web site is http://www.sdss.org/. The SDSS is
managed by the Astrophysical Research Consortium for the Participating
Institutions. The Participating Institutions are the American Museum
of Natural History, Astrophysical Institute Potsdam, University of
Basel, University of Cambridge, Case Western Reserve University,
University of Chicago, Drexel University, Fermilab, the Institute for
Advanced Study, the Japan Participation Group, Johns Hopkins
University, the Joint Institute for Nuclear Astrophysics, the Kavli
Institute for Particle Astrophysics and Cosmology, the Korean
Scientist Group, the Chinese Academy of Sciences (LAMOST), Los Alamos
National Laboratory, the Max-Planck Institute for Astronomy (MPIA),
the Max-Planck Institute for Astrophysics (MPA), New Mexico State
University, Ohio State University, University of Pittsburgh,
University of Portsmouth, Princeton University, the United States
Naval Observatory, and the University of Washington.

\bibliography{thebibliography}

\end{document}